\documentstyle[multicol,prl,aps,epsf,psfig,epsfig]{revtex}
\begin{document}
\title
{
Zero temperature  dynamics of Ising model on a densely connected 
small world network
}
\author
{
Pratap Kumar Das and Parongama Sen
}
\address
{
 Department of Physics, University of Calcutta,
    92 Acharya Prafulla Chandra Road, Kolkata 700009, India.
}
\maketitle
\begin{abstract}
The zero temperature quenching dynamics of the ferromagnetic Ising model 
on a densely connected small world network is studied where  long range 
bonds are added randomly with a finite probability $p$.  We find that in 
contrast to the sparsely 
connected networks and random graph,  there is no freezing and
an initial random configuration 
of the spins reaches the equilibrium
 configuration within a very few Monte Carlo time steps 
in the thermodynamic limit for any  $p \ne 0$. 
The residual energy and the number of spins flipped at any time shows an
exponential relaxation to equilibrium.
The persistence probability is also studied and it shows a saturation 
within a few time 
steps, the saturation value being 0.5 in the thermodynamic limit.
These results are explained in the light of the  topological  properties   
of the network which  is highly clustered and has a novel
small world behaviour.

\end{abstract}

  Reprint no: CU-physics-04/2005

\begin{multicols}{2}

\section{Introduction}
It is well known that the one dimensional Ising model with nearest neighbour interaction does
not have any non-trivial phase transition. 
However, a  drastic change is observed in its critical
behavior when even a few long
range interactions are added randomly. 
Such a  one dimensional lattice with extra random connections is
known to behave as a small-world network (SWN)\cite{Watts,newman,BA}, which  means that
 the  average shortest distance between any two sites in this lattice 
 scales with the 
logarithm of the number of
sites. The Ising model on  small world networks 
not only 
has a finite temperature phase transition \cite{Barrat,Gitter}, but the critical behavior is also 
mean-field like \cite{kim,herrero,hong,lopes,doro,leone}. A network in which the distribution of the number of links follows a power-law is known as a
scale free network (SFN). Here also a finite temperature phase transition 
of the Ising model with
 a diverging critical temperature \cite{scale} has been observed.
   
Recently, quite a few studies  
on the dynamical properties of the  Ising model 
on random graphs as well as networks have been reported both at
finite and zero temperatures \cite{sven1,hagg,boyer,sven2,zhu,jeong,castell}. 
 Dynamics of Ising models  is a much studied
phenomenon  and has emerged as a rich field of
present-day research. Models having same static critical behavior may display different
behavior when dynamic critical phenomena are considered \cite{Ho_Ha}. 
An important dynamical feature commonly studied is the quenching 
phenomenon below the critical temperature. 
In a quenching process, the system has a disordered initial configuration 
corresponding to a high temperature and its temperature is suddenly dropped.
This results in quite a few interesting phenomena like domain growth \cite{gunton,bray},
persistence \cite{satya1,derrida,stauffer,krap1,Krap_Redner} etc. 

In  one dimension, a zero temperature quench  of the  Ising model 
ultimately leads to the equilibrium
configuration, i.e., all spins point up (or down). 
 The average domain size $D$ increases in time $t$ as $D(t)\sim t^{1/z}$, 
where $z$ is the dynamical exponent. 
As the system coarsens, the magnetisation also 
grows in time as $m(t)\sim t^{1/2z}$. 
In two or higher dimensions, however, the system does not 
always reach equilibrium \cite{Krap_Redner} although these scaling relations
still hold good.

  Apart from the domain growth  phenomenon, another important dynamical 
  behavior commonly studied is persistence. In Ising model, persistence 
is simply the probability that a spin has not flipped till time $t$ and 
is given
by $P(t)\sim t^{-\theta}$. $\theta$ is called the persistence exponent and
is unrelated to any other static or dynamic exponents. Persistence  can also be 
studied at finite temperatures and the exponent may change at the critical 
temperature \cite{stauffer,derrida2}. 

 We have studied the zero temperature quenching dynamics of the ferromagnetic
Ising model on a SWN. It is important to carefully describe the type of network under consideration.
There are two well-known methods of generating  a small world network
starting from a chain of nodes having connections with nearest neighbours
only. These are (a) the addition type, where new long range (LR) bonds are added randomly
keeping the nearest neighbour connections intact, and (b)
the rewiring type where the existing nearest neighbour bonds
are rewired to distant neighbours randomly.
In the first case, when bonds are added with probability $p$,
the total number of LR bonds is $pN^2$ for large $N$ ($N$ is the number of nodes).
Even with $p \sim 1/N$, i.e., with a finite number of LR bonds, 
a phase transition
has been observed in the Ising model \cite{Gitter}. 
While considering the dynamics
of Ising models on addition type network, again $p = \gamma/N$ has been the usual
choice, where $\gamma $ is a finite quantity. The dynamical behaviour of
such a network for any value of $\gamma>1$ is comparable
to that of the random graph in the sense that the system fails
to reach its global minimum energy with zero temperature
Glauber dynamics even in the thermodynamic limit \cite{sven1,hagg}.
Quenching dynamics of the Ising model on the scale-free network
has also been considered recently \cite{castell} with an  average connectivity
$k=pN$ for each node. Here $k$ was fixed such that $p \sim 1/N$.
The results again show a freezing at zero temperature.

In our study, we have considered an addition type network generated from 
a one-dimensional chain of Ising spins with nearest neigbours.
Here each node has $pN$ number of LR bonds with $p$ fixed (i.e., finite in the limit $N \to \infty$) 
such that the
network is a densely connected network. In this network, we have
shown that the freezing or blocking effect is removed  for
any $p > 0$ in the thermodynamic limit. This conclusion is reached 
by studying various quantities like 
the  domain sizes,  magnetisation,
residual energy, number of flipped spins etc. as functions of time.
We also find that the relaxation to
the equilibrium state is exponential.

The study of persistence of the Ising model on this network 
shows that there is no algebraic decay as it reaches a constant finite value
similar to what happens in lattices of dimension greater than three
 \cite{stauffer}. The  behaviour of $P(t)$ 
with $p$  is described in detail later in the paper.

In section II we describe the dynamical model under consideration and 
the physical quantities calculated. The topological properties of the 
network are described in section III where we find that 
there is a novel behaviour of the network as far as 
small world property is concerened. The  results are discussed in section IV.
Summary and some conclusive comments have been given in section V.

\section{The Models and Dynamic properties studied}

    We have considered a one-dimensional ferromagnetic Ising model on a network,
 in which, apart from nearest neighbour links, 
there exist some random long range connections  with probability $p$.   
The Hamiltonian for this system is 
	\begin{equation}
	H=-\sum_{i}J_{ij}s_{i}s_{j},
	\end{equation}
where $s_i=\pm1$ is the state of the spin at the $i$th site, 
$J_{ij}=1 $ for nearest neighbours and
for other neighbours equal to 1 with probability $p$.
     The Hamiltonian  should be  divided by a factor of $pN$, 
     however, at $T=0$,
only the sign of the energy differences are required for the Glauber dynamics
and therefore this factor has not been included. 
 We have simulated this system with  periodic boundary condition. The initial 
configuration is random and single spin flip Glauber dynamics has been used for
subsequent updating, i.e., a spin is picked up at random and flipped if
the resulting configuration has lower energy, never flipped if
the energy is raised and flipped with probability 1/2 if there is 
no change in energy on flipping. \\  
  We have estimated the following quantities in the system.\\
$(1)$ Average domain size $D(t)$ at time $t$.\\
$(2)$ Magnetization $m(t)$ as a function of time (the average magnetisation
being zero from symmetry,   here $m(t)$ has been calculated by
taking the    average of  the 
absolute values of the magnetisation).\\
$(3)$ Residual energy $E_r(t) = E(t) -E_g$ where $E_g$ is the energy of the 
ground state and $E(t)$ the energy at time $t$.\\
$(4)$ $N_{flip}(t)$, the number of spin flips at time $t$ per unit time.\\ 
$(5)$ Persistence probability $P(t)$ defined as in section $I$.\\

When considering the domains, we have in mind  the original one-dimensional
lattice and measure the domain size along it.

  In finite dimensional nearest neighbour Ising models,   the dynamics of these quantities 
  is governed by the exponents $\theta$ and $z$, i.e.,\\
 $D(t)\sim t^{1/z}$       \\
 $m(t)\sim t^{1/{2z}}$  \\
 $E_r(t) \sim t^{-1/z}$\\
 $N_{flip}(t)\sim t^{-1/z}$\\  
 $P(t)\sim t^{-\theta}$   \\
 with $\theta=0.375$ and $z=2$ in one dimension.\\ 
In principle, it is not essential to study all these quantities to determine the characteristics of the dynamical system. However, in a numerical study, 
it is better to check that the behavior of these different quantities is 
consistent with a single $z$ and $\theta$ .

Since freezing is a key question here, we have also calculated the freezing 
probability $F(p)$ as a function of $p$ for various system sizes.
 Several other quantities and exponents related to domain dynamics in the  quenching phenomena can
 be defined \cite{krap2},
 but here  we have restricted our study to those mentioned above.

   In the simulations, we have restricted the system sizes to $N\leq 1000 $ as 
 a large number of configurations is required to have accurate data.
 The results have been averaged out over $1000$ initial
configurations and network configurations (typically).
As the network is densely connected, the updatings
consume a lot of CPU time forcing us to restrict our study to
rather small system sizes.

\section{Topological properties of the network}

The essential difference between a simple one dimensional 
 lattice with nearest neighbour links and the present network
lies in the topology of the networks. 
The topological properties of a network like the average  shortest distances 
 $\langle S \rangle$ (here distance means number of steps required to reach another node) and the clustering 
coefficient $\cal C$ may help in understanding the static and dynamical phenomena of Ising model on such network. 
In the nearest neighbour lattice, $\langle S \rangle$ is 
proportional to $N$  and 
$\cal C$ is zero (as there are no loops).

{\it Super small world effect}: The average shortest distance behaves 
in the expected manner; $\langle S\rangle$ is very small
($O(10)$) for very small values of $p$ and decreases to the exact value 1 in the
 $p\to 1$ limit. In a small world network, the shortest distance is 
supposed to scale as $\ln(N)$. 
Here, if $p$ is kept fixed, $\langle S\rangle$ actually decreases
with $N$ showing that it approaches  the behaviour corresponding to $p=1$
  in the thermodynamic 
limit. This is a novel behaviour  
 which one can call a  ``super small world'' effect as noted earlier in  \cite{zhou}. 
 One can also compare data for different $N's$ by 
keeping the number of edges per site, $pN$, a  constant rather than 
$p$. Therefore, in two networks 
of size $N_1$ and $N_2$, 
 the values of $p$ are kept $p_1$ and $p_2$ respectively 
such that $p_1N_1 = p_2N_2$. 
We  then observe that $\langle S_{N_1,p_1}\rangle /\langle S_{N_2,p_2}\rangle = \ln(N_1)/\ln(N_2)$
which is true for the conventional small world networks.
For example,  for  $N_1=100$ and  $p_1=0.05$, ~~$\langle S_{100,p_1}\rangle=2.58 $
and for $N_2=500$ and  $p_2=0.01, ~~\langle S_{500,0.01}\rangle=3.43$, and the ratio of these
two quantities is very close to that of $\ln(100)$ and $\ln(500)$.

The clustering coefficient is estimated in the standard way: 
the probability that  two neighbours of a particular node
are also  connected to each other is a measure of the   clustering tendency.
$\cal C$ is also calculated for different $p$ and $N$ values.
Obviously it increases with $p$ and is equal to one at $p=1$
when all the nodes are connected to each other. Since the nearest neighbours
are already connected, clustering coefficient would increase
if the long range bonds happen to be  next nearest neighbours \cite{PS}. For 
small values of $p$, this is more probable in
smaller lattices and therefore  clustering coefficient
shows a marked decrease with $N$.
Fig. $1$ shows the data for both $\langle S \rangle$ and $\cal C$.

\begin{center}
\begin{figure}
\noindent \includegraphics[clip,width= 6cm, angle=270]{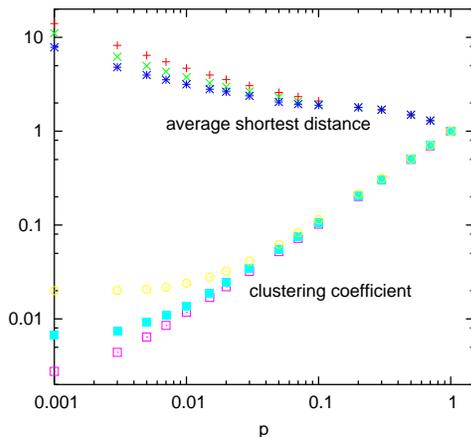}

\caption{Average shortest distance $\langle S \rangle$ and clustering 
coefficient $\cal C$ as functions of $p$ for different system sizes.
In both cases, the values are lesser for larger $N$ values.}
\end{figure}
\end{center}

These results will be helpful to interpret the observations
which we have made and will be referred to in the later sections.

\section{RESULTS and DISCUSSIONS}

We first discuss the results for the growth of the  domains sizes and magnetisation.
The domain sizes have been scaled  by the system sizes $N$ such that $D(t) \leq 1$. 
We have verified that   for $p=0$,
all the physical quantities under consideration  follow the 
known behaviour summarised in section II (with $z$ and $\theta$ assuming the
values corresponding to one dimension).
 As soon as a non-zero $p$ 
value is introduced, both $m(t)$ and $D(t)$ quickly 
reach a saturation value  such that  a variation with time occurs only over a 
short
initial period of time (Fig. 2). 

\begin{center}
\begin{figure}
\noindent \includegraphics[clip,width= 6cm, angle=270]{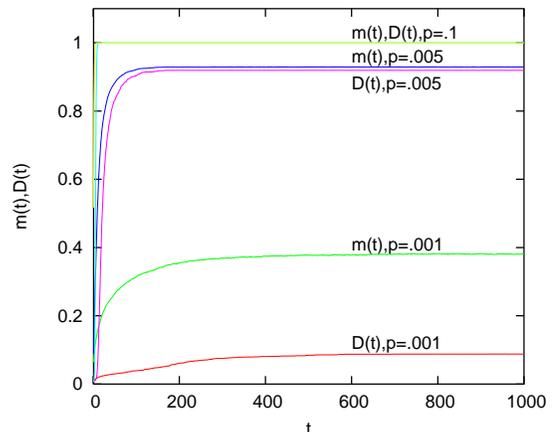}
\caption{Magnetisation $m(t)$ and domain size $D(t)$  as functions of time $t$ for
different values of $p$ with the system size $N = 500$. Both 
saturate very quickly as $p$ is increased.}
\end{figure}
\end{center}

For small system sizes and at low $p$, 
the saturation values of $m(t)$ and $D(t)$ are far from those of  the 
equilibrium values (both unity at $T$=0) similar to the results on random
graphs and sparsely connected small world networks. 
This apparently suggests that the system gets ``frozen''
in one of the metastable states. 
However, in the present model, blocking seems to be effective only for 
finite sizes, as  the saturation values of both magnetisation and average
domain size approach unity (i.e., the equilibrium value) when the system
size is increased (Fig. 3). This is true for any finite value of $p \neq 0$.
The apparent blocking effect is  more prominent for small values
of $p$.

\begin{center}
\begin{figure}
\noindent \includegraphics[clip,width= 6cm, angle=270]{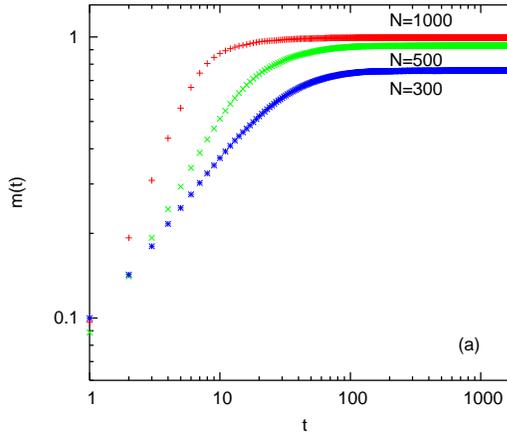}

\caption{Magnetization $m(t)$  as function of $t$ 
for different system sizes at $p=.005$ is shown. Saturation value is higher 
for larger size at a particular $p$ implying that blocking is a finite 
size effect here. This behaviour is true for all $p>0$. The domain 
size $D(t)$ shows similar behaviour.}
\end{figure}
\end{center}

 The comparison of  $m(t)$ and $D(t)$
for different system sizes also shows that the 
initial  growth becomes rapidly sharper with the system size, so that
any time dependence in the initial period loses its significance.
The period over which this growth takes place also shrinks in size
in larger system sizes (Fig. 3) signifying a very fast growth in the magnetization.

The study of the distribution of magnetisation
is also consistent with the fact that blocking occurs for finite sizes
only. The distribution has finite values for
all values of $m$ ($ -1 < m < 1$) for small $N$,
but for larger sizes has non-zero values very close to $m =\pm 1$ only.

\begin{center}
\begin{figure}
\noindent \includegraphics[clip,width= 6cm, angle=270]{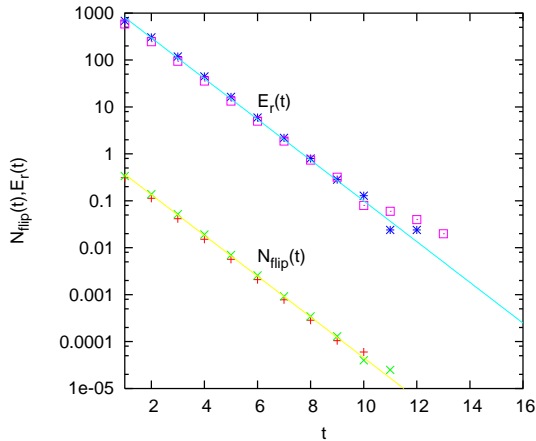}
\caption{$N_{flip}(t)$ and scaled residual energy $E_r(t)/p$ as function of $t$ 
for two different values of $p$ ($p=.1$ and $.8$) with the system 
size $N = 1000$. Both $N_{flip}(t)$ and
$E_r(t)$ follow exponential decay as $\exp(-\alpha t)$ with $\alpha=1$ 
for all $p$ and all system sizes. The straight lines have slope $=1$.}
\end{figure}
\end{center}

The fast growth in  $m(t)$ and $D(t)$ is supported by the behaviour of
$E_r(t)$ and  $N_{flip}(t)$ as both show an exponential decay ($E_r(t), ~N_{flip}(t) \sim
\exp(-\alpha t)$) with $\alpha =1$ for all $p$ (Fig. 4).
\medskip

{\it{Freezing probability}} 

\medskip

The above observations suggest that for finite sizes, the probability
that the system goes to a frozen state is finite for small values of
$p$. Since for the limiting cases $p=1$ and $p=0$, there 
is no freezing, it is expected that the freezing probability $F(p)$
will have a peak for a non-zero value of $p$. We study the freezing
probability for fixed values of $N$ and find out some interesting features.
There is indeed a peak occurring at $p=p_m$   with $p_m \sim 1/N $.
Except for very small values of $p$, $F(p)$ decreases with $N$ signifying the
disappearance of freezing in the thermodynamic limit. For small values of $p$,
 the behaviour may be different. 
Specifically, if   $p = 1/N$,  we  find that the freezing probability
$increases$ with $N$, which is in consistency with the observation
of \cite{castell}. But at this value of $p$,
the network is a sparsely connected one and 
therefore  it has a freezing tendency indicated by this increase.
In fact, keeping $p$ fixed at a certain value such that 
$p < p_m$ (for the system sizes concerned) we find  that the  freezing tendency 
gets enhanced with $N$.
Also, we find that $F(p)$ shows an exponential decay beyond $p_m$: $F(p) \sim \exp(-\gamma p)$
for large values of $p$ with $\gamma \sim N$.

\begin{center}
\begin{figure}
\noindent \includegraphics[clip,width= 8.5cm, angle=270]{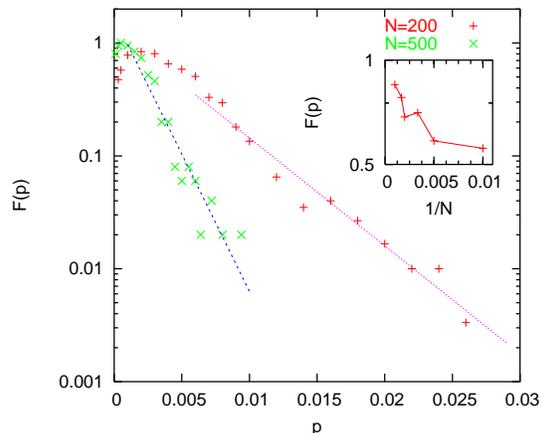}
\caption{ Freezing probability $F(p)$ as a function of $p$
for different system sizes. 
$F(p)$  has a peak which shifts towards zero for  larger $N$. 
The tail of $F(p)$ follows exponential decay 
as $\exp(-\gamma p)$ with $\gamma \sim N $ shown by the dotted lines.
Inset shows $F(p)$ against $1/N$ for $p=1/N $.}
\end{figure}
\end{center}

These results (shown in Fig. 5) indicate again that there is a sharp discontinuity
in the dynamical behaviour at $p=0$ and that freezing disappears for any non-zero $p$ for
large values of $N$
as long as $p$ is a finite quantity.

In reference \cite{boyer}, it had been shown how the domain walls get pinned
when the number of extra bonds is $ O(N) $, i.e., $p\sim 1/N$ and the
system freezes. 
The domain walls become mobile as soon as the number of long 
range interactions increase to $O(N^2)$ and 
ultimately they disappear by annihilating each other. Thus, in a densely 
connected infinite network, freezing disappears for any $p\neq 0$. 

We next study the behaviour of the persistent probability $P(t)$
with time for different values of $p$ (Fig. 6).
$P(t)$ follows the
well known power law decay for $p=0$, but quickly falls to a finite
saturation value $P_{sat}$ for any non-zero value of $p$. The decay is 
sharper for higher values of $p$.
The saturation behaviour is similar to that of $D(t)$ and $m(t)$.

The behaviour of $P_{sat}$ with different system sizes however,  shows an 
interesting feature.
In a finite $d$-dimensional lattice   of  size $L$,   when  $P(t)$ decays algebraically in time with 
the exponent
$\theta$,  the persistent probability as a 
function of $t$ and $L$ is given by
 
\begin{equation}
P(L, t) \sim t^{-\theta}f(t/L^z),
\end{equation}
where $f(x) =$ constant for small $x$ and $f(x) =x^\theta$ for large $x$ 
such that $P(L, t\to \infty) \sim L^{-z\theta}$ which indicates 
that the time independent
persistence probability decreases with $N$, where $N=L^d$, the total number
of lattice sites. For non-zero values of $p$, when
there is no algebraic decay of the persistent probability with time,
we find that (Fig. 7) there exists a value of $p=p^*$ below which 
the persistence 
probability actually increases with $N$, the increase with $N$ being slower 
than a power law.
Beyond $p^*$, $ P_{sat}$ decreases with  $N$ but the decrease is fairly weak.
However for both  $p < p^*$ and $p > p^*$, $P(N,t,p)$ approaches a constant 
close to 0.5 from below and above respectively.
That the saturation value of the persistence probability is close to
0.5 can be justified: initially fifty persent spins are up/down,
and  spins of only one type are  flipped only within the short time the 
system reaches the equilibrium state.
The value of $p^* \simeq 0.25$. 

\begin{center}
\begin{figure}
\noindent \includegraphics[clip,width= 6cm, angle=270]{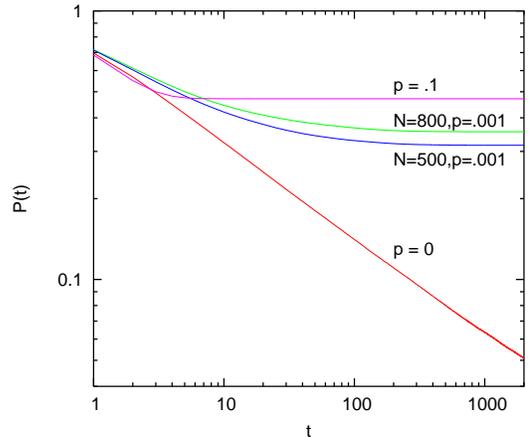}
\caption{Persistence $P(t)$ as a function of time $t$ for different probability
$p$ for system size $N=500$. $P(t)$ follows the well known power law decay
as $t^{-\theta}$ with $\theta=0.375$ at $p = 0$. At finite $p$, $P(t)$ decays to
a constant value within a few time steps. For higher $p$, the decay is even
faster and the dynamics stops earlier.}
\end{figure}
\end{center}

\begin{center}
\begin{figure}
\noindent \includegraphics[clip,width= 6cm, angle=270]{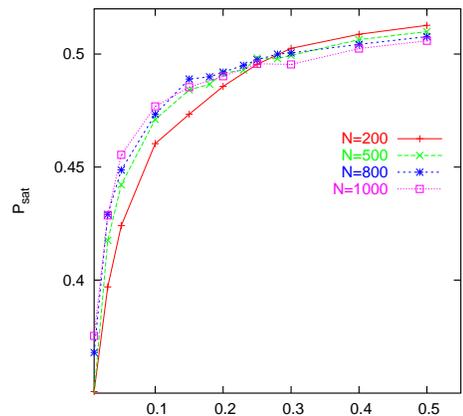}
\caption{Saturation value of $P(t)$ as a function of $p$ for different system sizes.}
\end{figure}
\end{center}

\section{Summary and Conclusions}

We have studied the dynamics of the ferromagnetic Ising model on a small-world
network at zero temperature. The network is densely connected in the
sense that there is a finite number of extra bonds for each 
node. It is observed that addition of long range bonds with probability $p$
brings the initial random configuration of Ising spins to the equilibrium 
ferromagnetic configuration within a very few time steps even if $p$ is very
small. Magnetisation and average domain size  quickly reach  the 
saturation values without showing any scaling behaviour
with time. Consistent with this observation, 
the residual energy and the number of flippings at any time  shows an 
exponential decay.  
The study of the freezing probability shows a peak occurring 
in the distribution at $p_m \sim 1/N$ and an exponential decay which becomes
faster with the system size.
Persistence probability  also reaches a saturation value within
a few time steps. 
These  results obtained for the quenching dynamics 
indicate that the dynamical behaviour of this densely connected
network is much
different  compared to that of random graphs and sparsely connected SWN
 where the same dynamics  leads to a frozen state not
equivalent to the equilibrium ground state.
That there is no power law behaviour but exponential relaxation is
consistent with the mean field behaviour of the network \cite{silva}.

In order to explain our observations, we notice that  
the present model is highly clustered as 
the density of connections is large (see section III). In contrast, the addition
type small world network (with two nearest neighbour only),
generated with a $p \sim 1/N$ has a vanishing clustering coefficient
even though it has small world property \cite{PS}. This is in fact the reason
for its behaviour as a random graph which  also has a small world
effect but vanishing clustering coefficient.
The large clustering in our model is effective in 
making the domains entirely non-local and therefore 
 the system can reach the global equilibrium state very fast.
The super small world effect, by which we mean that the average 
shortest distance  decreases with $N$ (tending to unity),
also helps in understanding the results. For any $p \neq 0$, the network
flows towards  the $p=1$ fixed point (for large sized networks) 
for which one does not expect any freezing.

Our results are consistent with some   very recent 
observations \cite{castell}
where the dynamics of Ising model on a scale free network with increasing number
of connectivity $k=pN$ has been studied.  From the data shown,  
one can obtain the freezing probability
as a function of a fixed $p$ (e.g., $p=0.1$) as well, 
which shows that freezing will $disappear$ for large system sizes.

Usually persistence probability has a unique behaviour and is governed by an
independent exponent $\theta$ not related to the dynamic
exponent $z$ which dictates the behaviour of $D(t), ~m(t), E_r(t)$ and $N_{flip}(t)$. 
Here one cannot make any statement about the independence of 
persistence and the domain growth phenomena.
The only distinctive behaviour of persistence
seems to be a difference in  behaviour with finite sizes occurring for $p< p^*$
where $p^* \simeq 0.25$. 
At present our understanding of the network and the dynamics is not enough
to explain the significance of $p^*$, although the limiting value of the
persistence probability being 0.5 is easily explained from the mean
field nature of the network.

From the results reported in the present paper, we conclude that for
quenching dynamics, 
for any $p \le 1/N$ the time evolution leads to a frozen state
far from equilibrium whereas with a finite $p$ freezing is overcome.
The finite density of connections thus acts as a driving force, like a finite
temperature, which drives the system out of the frozen state. This may appear
as a dynamical phase transition in finite systems, 
however, in the thermodynamic limit ($N \to \infty$) of course, there is no such
transition.

We finally remark that as far as statics is concerned, a finite value
of
$p$ is not required to get a phase transition but the  finiteness  
of $p$   is essential to remove the dynamic frustration
when dynamics is considered.

%
\vskip .5cm

Acknowledgments: We thank 
S. Dasgupta and S. N. Majumdar
for helpful discussions.
P. K. Das acknowledges support from CSIR grants no.
9/28(608)/2003-EMR-I.
 PS acknowledges DST grant number SP/S2/M-11/99.

\vskip .8cm
Email:  {pra$\_$tapdas}@rediffmail.com, parongama@vsnl.net

\end{multicols}

\end{document}